\documentclass{article}
\usepackage{spconf,amsmath,graphicx}
\usepackage{amsmath}
\usepackage{amssymb}
\usepackage{amsfonts}
\usepackage{amstext}
\usepackage{amsthm}
\usepackage{algorithm} 
\usepackage{algpseudocode}
\usepackage{float}% http://ctan.org/pkg/float
\usepackage{booktabs}
\usepackage[hidelinks]{hyperref}
\usepackage[title]{appendix}

\newtheorem{theorem}{Theorem}

\title{Using fine-tuning and Min Lookahead beam search to improve Whisper}

\name{Andrea Do$^{\star}$ \enspace Oscar Brown$^{\dagger \star}$ \enspace Zhengjie Wang$^{\star}$ \enspace Nikhil Mathew$^{\star}$ \enspace Zixin Liu$^{\star}$ \enspace Jawwad Ahmed$^{\star}$ \enspace Cheng Yu$^{\star}$\thanks{This work was funded by Trellis Data Group. ML Research Labs is a subsidiary of Trellis Data Group.}}
\address{ML Research Labs$^{\star}$, Australian National University$^{\dagger}$}

\begin{document}
\ninept

\maketitle

\begin{abstract}
The performance of Whisper in low-resource languages is still far from perfect. In addition to a lack of training data on low-resource languages, we identify some limitations in the beam search algorithm used in Whisper. To address these issues, we fine-tune Whisper on additional data and propose an improved decoding algorithm. On the Vietnamese language, fine-tuning Whisper-Tiny with LoRA leads to an improvement of 38.49 in WER over the zero-shot Whisper-Tiny setting which is a further reduction of 1.45 compared to full-parameter fine-tuning. Additionally, by using Filter-Ends and Min Lookahead decoding algorithms, the WER reduces by 2.26 on average over a range of languages compared to standard beam search. These results generalise to larger Whisper model sizes. We also prove a theorem that Min Lookahead outperforms the standard beam search algorithm used in Whisper.

\end{abstract}
\section{Introduction}
\label{sec:intro}

    Whisper has remarkable performance in transcribing multilingual speech audio into text \cite{radford2022robust}. While its performance with English and other high-resource languages is impressive, the limited availability of training audio data for low-resources languages is a challenge. As Whisper is open-source, researchers may enhance its performance with new training datasets and methods. In this paper, we investigate unconventional fine-tuning and decoding algorithms to improve Whisper's performance in a low-resource scenario.

    While fine-tuning is common in practice, a systematic comparison between different fine-tuning strategies for an encoder-decoder model like Whisper has yet to be documented. In the work of Jain \textit{et al.} \cite{jain2023adaptation}, the authors froze most of the model's parameters while fine-tuned only the final layer. Conversely, Rouditchenko \textit{et al.} \cite{rouditchenko2023comparison} fine-tuned the entire model on unseen languages. Both studies lack comprehensive explanations for their choice of fine-tuning strategies. To fill this gap, we conduct a comprehensive study of fine-tuning strategies on Whisper, including full-parameter fine-tuning and partial-parameter fine-tuning where gradients are updated only in parts of the model. We selected Vietnamese as our target language, but we believe the results translate to other low-resource languages since we did not utilise any language specific features in our fine-tuning experiments. 

    Whisper uses a beam search decoding algorithm with beam width $n=5$ and log-probability (logprob) as the score function \cite{radford2022robust}. This is as opposed to the greedy algorithm which chooses the token with the greatest logprob at each decoding step. Although beam search outperforms the greedy algorithm, we suggest it can be further improved by filtering out certain sequences and performing additional decoding to access the probabilities of tokens at later decoding steps.

    The rest of this paper is structured as follows: Section 2 details the methodology used in fine-tuning and the decoding algorithm. Section 3 documents the various experiments we conducted to improve Whisper models. Finally, we conclude with Section 4 where we discuss the contribution of our paper and outline the future work.
\section{Methods}
\label{sec:methods}
\subsection{Whisper Model Architecture}
    Whisper employs a straightforward encoder-decoder Transformer architecture \cite{vaswani2023attention}. The encoder takes a log-magnitude Mel spectrogram $M$ and outputs a sequence of embeddings
    \begin{equation}
        H = (h_1, \ldots, h_{T^{\prime}})=f_{\theta_{\textrm{encoder}}}(\textrm{M}).
    \end{equation}
    The input to the decoder includes a set of Whisper multitask special tokens $v_S$ \cite{radford2022robust}, the output from the encoder and the current output text tokens $(v_0, \ldots, v_{i}$). At each step, the decoder produces logits for each token in a vocabulary
    \begin{align}
        l_{i+1} &= f_{\theta_{\textrm{decoder}}}(v_S, H, v_0, \ldots, v_i),
    \end{align}
    which is then generated into an output sequence.
    The audio encoder has two convolution layers, a token embedding layer and a block of Encoder Transformer layers. The text decoder has a trainable token embedding and a number of Decoder Transformer layers. Radford \textit{et al.} \cite{radford2022robust} trained Whisper models of various sizes, only differing in the number of Transformer layers.
    % The architecture of the Transformer layers in both are similar, with input passed through a self-attention layer followed by a feed-forward layer. However, the Transformer layers in the decoder have an additional cross-attention layer in between its self-attention and feed-forward layer, where the attention matrix between the output of the decoder self-attention layer and the output of the encoder is computed.
\subsection{Fine-tuning}
\subsubsection{Decoupling token input and output embeddings}
    Whisper uses tied input-output token representations which was introduced in \cite{press2017using}, whose LSTM translation model experiments showed an advantage in reducing the model size without compromising model performance. However, when revisited in Transformer-based models \cite{chung2020rethinking}, decoupling the input and output token embeddings enhanced the generality and transferability of large pretrained language models upon increasing output embedding size. This suggests it may be beneficial to let the model learn input and output token representations in isolation. In Whisper, the final logit is a product of the output of the last attention layer $x$ and the token embedding $E$, 
    \begin{equation}
        l_i = x E^{\text{T}}.
    \label{equation:decoder-logit}
    \end{equation}
    In this study we investigated decoupling embedding layers by replacing $E$ in Equation (\ref{equation:decoder-logit}) with an output embedding $E_{\text{out}}$ whose dimensions are similar to $E$. We initialised $E_{\text{out}}$ as an embedding layer and copied the weight from $E$ in two experiments: fine-tuning the entire model with decoupling and fine-tuning only the decoder embedding with decoupling, Table \ref{table:finetuning-tiny}.
    \begin{table}[h]
    \centering
    \caption{\textit{WER (\%) results on FLEURS and CommonVoice 9 with different fine-tuning strategies on Whisper-Tiny after training on 100 hours of Vietnamese speech data prior to greedy decoding}. }
    \begin{tabular}{lccc} 
         \toprule
         Fine-tune strategy &
         \begin{tabular}{@{}c@{}}Trainable \\ Params (M)\end{tabular}
         & FLEURS &
         \begin{tabular}{@{}c@{}}Common \\ Voice 9\end{tabular}
         \\ [0.5ex] 
         \midrule
         Zero-Shot & & 74.78 & 91.44\\
         Full-parameter & 37.20 & 37.74 & 35.36\\ 
         \begin{tabular}{@{}l@{}}Full-parameter \\ w/ decoupling\end{tabular}
         & 57.10 & 36.60 & \textbf{34.14} \\
         Decoder only & 29.55 & 42.74 & 48.13\\
         Decoder embed. & 20.00 & 54.12 & 56.83\\
         \begin{tabular}{@{}l@{}}Decoder embed. \\ w/ decoupling\end{tabular} & 40.00 & 51.90 & \ 53.77\\
         LoRA\\
         $r = 192$, $\alpha = 384$ & 23.00 & \textbf{36.29} & 35.51\\
         $r = 96$, $\alpha = 192$ & 11.50 & 37.07 & 36.40\\
         $r = 32$, $\alpha = 64$ & 3.83 & 39.07 & 35.70\\
         \bottomrule
    \end{tabular}
    \label{table:finetuning-tiny}
    \end{table}
\subsubsection{LoRA}
    To address the issue of fine-tuning large models with limited resources, the research community has directed its focus towards parameter-efficient fine-tuning techniques such as LoRA \cite{hu2021lora} and Adapters \cite{houlsby2019parameterefficient}. Among these approaches, LoRA has gained considerable interest due to its comparable performance to full-parameter fine-tuning, while utilising only a fraction of the compute power \cite {sun2023comparative}. LoRA can also be integrated into a range of Transformer-based models. Hu \textit{et al.} \cite{hu2021lora} applied a low-rank decomposition to the updates of a pre-trained weight matrix $W_0 \in \mathbb{R}^{d \times k}$ with
    \begin{equation}
        W_0 + \Delta W = W_0 + BA,
    \end{equation}
    where $B \in \mathbb{R}^{d \times r}$ and $A \in \mathbb{R}^{r \times k}$ are smaller matrices of rank $r \ll \min(d, k)$. During fine-tuning, only $A$ and $B$ receive gradient updates while $W_0$ is frozen. $\Delta W$ is further scaled by a fraction $r/\alpha$. Thus, LoRA is characterised by two hyper-parameters: the rank $r$ and the scaling $\alpha$. 
    In this paper, we focus on using LoRA modules to fine-tune Whisper on a low-resource language in which the training objective remains unchanged. We implemented the original LoRA method together with the decoupling of token embeddings. Here, we initialised $E_{\text{out}}$ as a linear layer whose weights were copied from $E$ and bias was 0. Then drawing inspiration from \cite{xu2023baize,sun2023comparative,zhang2023adaptive}, we extended the use of LoRA modules to all linear layers within the Whisper model, including the query, key, value, output projection, MLP, and $E_{\text{out}}$. 
\subsection{Decoding algorithms}
\subsubsection{Filter-Ends}
    Whisper's decoding algorithm stores a beam containing the score of $n$ sequences of tokens of length \(k\) and generates from these the $n$ sequences of tokens of length \(k+1\) with the greatest sum of logprobs of tokens. This continues until $n$ sequences ending in an End Of Transcript (EOT) token have occurred in the beam. It then chooses the sequence with the greatest sum of logprobs, dividing by the length of the sequence for normalisation \cite{radford2022robust}. This normalisation often creates contradictions with the probabilities that the model assigns. One notable example is when sequence \(A\) of length \(n\) has the same tokens as sequence \(B\) before position \(n\), but \(A\) has an EOT token which the decoder assigns higher logprob than the \(n\)-th token of \(B\). It is possible for \(B\) to be chosen if the later tokens in \(B\) have sufficiently high logprobs. However, this contradicts the model's prediction that \(A\) is more likely than any sequence beginning with the first \(n\) tokens of \(B\). To correct this, we devise a modification called Filter-Ends (FE), which filters out every token less likely than an EOT token given the same base sequence.

    \begin{algorithm}[h]
	\caption{Min Lookahead}
	\begin{algorithmic}[1]
            \Require a beam width $n$, a lookahead $m$ and elements of the beam $b_1,b_2,\ldots,b_n$ where each $b_i$ is a sequence of tokens
            \Ensure $S_{\textrm{best}}$, beam for the next step of decoding
            \State Let $S_{\textrm{potential}}$ be an empty list.
            \ForAll {$b_i$ in $b_1,b_2,\ldots, b_n$}
                \State Run decoder on $b_i$ to get probabilities of next tokens
                \State Concatenate $b_i$ with the $n$ most likely next tokens
                \State Append these $n$ concatenations to $S_{\textrm{potential}}$
            \EndFor
            \ForAll {$s_i$ in $S_{\textrm{potential}}$}
                \State Let $q_{i,0}$ be the probability of $s_i$
                \State $s_{\textrm{current}}=s_i$
    		\For {step $= 1$ to $m$}
    			\State Run decoder on $s_{\textrm{current}}$
                    \State Concatenate the most likely token to $s_{\textrm{current}}$
                    \State Let \(p_1,...,p_n\) be the probabilities in descending order of top $n$ most likely tokens
                    \State \(t_{i,\textrm{step}}=\dfrac{p_1\log(p_1)+\ldots+p_n\log(p_n)}{p_1+\ldots+p_n}\)
                    \State \(q_{i,\textrm{step}}=\dfrac{p_1}{p_1+\ldots +p_n}q_{i,\textrm{step}-1}\)
    		\EndFor
            \EndFor
            \State Let $S_{\textrm{best}}$ be an empty list
            \ForAll {$s_i$ in $S_{\textrm{potential}}$}
                \ForAll {sequences $s_j$ in $S_{\textrm{best}}$}
                    \If {\(\sum_{k=1}^m ((t_{i,k}-t_{j,k})\min(q_{i,k-1},q_{j,k-1})) + \log(q_{i,0})-\log(q_{j,0}) > 0\)}
                        \State Insert $s_i$ in $S_{\textrm{best}}$ immediately before $s_j$
                        \State Remove the last element of $S_{\textrm{best}}$ if $S_{\textrm{best}}$ has more than \(n\) elements
                    \EndIf
                \EndFor
                \State Append $s_i$ to $S_{\textrm{best}}$ if $S_{\textrm{best}}$ has less than \(n\) elements
            \EndFor
	\end{algorithmic}
    \label{algorithm:min-lookahead}
    \end{algorithm}
\begin{table*}[h]
    \centering
    \caption{\textit{WER (\%) results on the FLEURS test set for different models and decoding algorithms. Ours is Filter Ends + Min Lookahead = 3.}}
    \begin{tabular}{l *{12}c}
        \toprule
         & \textbf{Eng\textsuperscript{1}} & \textbf{Spa\textsuperscript{1}} & \textbf{Fra\textsuperscript{2}} & \textbf{Pol\textsuperscript{2}} & \textbf{Tur\textsuperscript{2}} & \textbf{Ind\textsuperscript{2}} & \textbf{Vie\textsuperscript{3}} & \textbf{Hun\textsuperscript{3}} & \textbf{Lit\textsuperscript{4}} & \textbf{Hye\textsuperscript{4}}  & \textbf{Kan\textsuperscript{4}} & \textbf{Improvement} \\
        \midrule
        \textbf{Tiny} \\
        OpenAI Value \cite{radford2022robust} & 12.4 & 15.9 & 41.4 & 45.6 & 42.5 & 51.7& 60.0 & 83.8 & - & - & - & -0.01 \\
        Beam Search & 11.81 & 15.70 & 37.71 & 48.45 & 40.32 & 51.73 & 58.17 & 89.36 & - & - & - & 0.00 \\
        Filter-Ends & 11.86 & 15.70 & 37.56 & 46.94 & 40.07 & 51.40 & 57.91 & 84.31 & - & - & - & 0.94 \\
        Ours & \textbf{11.39} & \textbf{15.24} & \textbf{36.66} &\textbf{44.46} & \textbf{38.39} & \textbf{50.20} & \textbf{56.74} & \textbf{82.07} & - & - & - & \textbf{2.26} \\
        \midrule
        \textbf{Base} \\
        OpenAI Value \cite{radford2022robust} & 8.9 & 9.9 & 28.5 & 33.1 & 30.8 & 27.5 & 40.5 & 65.0 & 87.3 & - & - & 1.25 \\
        Beam Search & 8.56 & 10.02 & 24.00 & 32.70 & 27.63 & 34.15 & 41.00 & 73.65 & 91.05 & - & - & 0.00 \\
        Filter-Ends & 8.53 &  9.99 & 23.98 & 31.73 & 26.75 & 33.70 & 40.15 & 65.65 & 88.05 & - & - & 1.58 \\
        Ours & \textbf{8.45} & \textbf{9.80} & \textbf{23.43} &\textbf{30.77} & \textbf{25.97} & \textbf{32.10} & \textbf{39.24} & \textbf{63.87} & \textbf{86.91} & - & - & \textbf{2.47} \\
        \midrule
        \textbf{Small} \\
        OpenAI Value \cite{radford2022robust} & 6.1 & 5.6 & 15.0 & \textbf{14.7} & 15.9 & 16.3 & \textbf{21.2} & 38.9 & \textbf{65.6} & 86.6 & - & -0.08 \\
        Beam Search & 5.94 & 5.49 & 12.62 & 15.34 & 14.52 & 16.53 & 21.96 & 40.09 & 67.59 & 85.00 & - & 0.00 \\
        Filter-Ends & 5.94 & 5.48 & 12.64 & 15.34 & 14.59 & 16.53 & 21.43 & 39.12 & 66.72 & 83.12 & - & 0.42 \\
        Ours & \textbf{5.83} & \textbf{5.45} & \textbf{12.30} & 15.11 & \textbf{14.36} & \textbf{16.12} & 21.25 & \textbf{38.55} & 65.69 & \textbf{82.70} & - & \textbf{0.77} \\
        \midrule
        \textbf{Medium} \\
        OpenAI Value \cite{radford2022robust} & 4.4 & \textbf{3.6} & 8.7 & \textbf{8.0} & 10.4 & \textbf{10.2} & \textbf{12.7} & \textbf{24.3} & \textbf{41.1} & 60.1 & 77.7 & -1.00 \\
        Beam Search & \textbf{4.35} & 3.85 & 7.74 & 8.20 & \textbf{8.77} & 10.41 & 13.11 & 26.53 & 43.63 & 56.91 & 66.73 & 0.00 \\
        Filter-Ends & \textbf{4.35} & 3.86 & 7.73 & 8.28 & 8.80 & 10.43 & 13.10 & 26.51 & 42.32 & 56.53 & \textbf{66.52} & 0.16 \\
        Ours & 4.47 & 3.82 & \textbf{7.71} & 8.23 & 8.88 & 10.28 & 12.91 & 26.05 & 41.71 & \textbf{56.29} & 67.04 & \textbf{0.26} \\
        \midrule
        \textbf{Large-v2} \\
        OpenAI Value \cite{radford2022robust} & 4.2 & \textbf{3.0} & 8.3 & \textbf{5.4} & 8.4 & 7.1 & \textbf{10.3} & 17.0 & \textbf{28.1} & 44.6 & \textbf{37.0} & 0.40 \\
        Beam Search & \textbf{4.06} & 3.20 & 5.69 & 5.54 & 7.34 & 7.17 & 11.09 & 17.25 & 31.31 & 44.82 & 40.27 & 0.00 \\
        Filter-Ends & \textbf{4.06} & 3.21 & \textbf{5.67} & 5.56 & 7.35 & 7.17 & 10.51 & 17.28 & 29.27 & 44.59 & 40.05 & 0.28 \\
        Ours & 4.07 & 3.20 & 5.69 & 5.46 & \textbf{7.29} & \textbf{7.10} & 10.55 & \textbf{16.90} & 28.94 & \textbf{44.26} & 39.42 & \textbf{0.44} \\
        \bottomrule
        \multicolumn{13}{l}{\textsuperscript{1}\footnotesize{Language with $> 10000$ hours of speech data }} {\textsuperscript{2}\footnotesize{Language with $1000$-$10000$ hours of speech data }}
        {\textsuperscript{3}\footnotesize{Language with $100$-$1000$ hours of speech data }} \\
        \multicolumn{13}{l}{\textsuperscript{4}\footnotesize{Language with $< 100$ hours of speech data }}
    \end{tabular}
    \label{table:decoding-algorithm-all}
    \end{table*}
\subsubsection{Min Lookahead in Beam Search}
    Zheng \textit{et al.} \cite{zheng2019speculative} suggests an improvement to the basic beam search algorithm where in the sum of logprobs of tokens, the logprobs for the most likely future token in the next \(m\) steps given the chosen tokens, are also included. We call this algorithm `basic lookahead'. We find the following flaws in this approach: the future tokens are not guaranteed to be what the beam search algorithm would choose since they are chosen through greedy decoding, and the logprobs of the next most likely future tokens chosen by the decoder are not considered at all. We devise a new lookahead algorithm that we call `min lookahead'. In Appendix \ref{appendix: min lookahead proof}, we prove that this algorithm is expected to outperform at standard beam search. See Algorithm \ref{algorithm:min-lookahead} for details of the approach.
    
    The motivation of the algorithm is that we weight the logprob score function of a possible future token by the probability that the token is used. Additionally, we do this for the top \(n\) tokens at each lookahead step rather than the top \(1\), since the decoder already outputs the probabilities of these. When comparing the token in the same lookahead step and likelihood rank for two potential sequences, we take the minimum of the probabilities of each token occurring so that both sums have the same total weighting and no terms are overly weighted.
    We have also tested this algorithm using maximum and mean functions instead of a minimum function on line 22 in Algorithm \ref{algorithm:min-lookahead}. Although, the maximum variation is not necessarily expected to outperform a standard beam search, there is a very similar proof to Appendix \ref{appendix: min lookahead proof} that the maximum variation outperforms basic lookahead.
\section{Experiments}
\label{sec:experiments}
    
\subsection{Fine-tuning Whisper-Tiny}
    All fine-tuning experiments were conducted using an A40 48GB GPU, with a batch size of 32 and a learning rate of $10^{-5}$ \cite{rouditchenko2023comparison}. Unlike previous studies which focused on training efficiency, we pushed the limits of LoRA by setting $r = 192$, which is half the number of attention dimensions in Whisper-Tiny. 
    For training, we used the VinBigdata-VLSP2020-100h dataset \cite{vlspResourcesAssociation}, a 100-hour Vietnamese ASR dataset published by VinBigData in 2020 to simulate a realistic scenario for fine-tuning Whisper models on low-resource languages. This includes 20 hours of read speech sourced from speakers reading manually prepared transcripts, and 80 hours of web-crawled speech manually transcribed with over 95\% accuracy. 
    An existing dataset was used since it is difficult to create a high-quality ASR dataset from scratch, owing to the complexity of validation without mastery over the language. We employed a native Vietnamese speaker to meticulously sample and verify the dataset prior to utilisation. We understand that this dataset lacks benchmarks or prior publications. To select the optimal checkpoint and assess performance, we used the FLEURS \cite{conneau2022fleurs} validation set and reported results on both the FLEURS and CommonVoice 9 \cite{commonvoice:2020} test sets. 
    Table \ref{table:finetuning-tiny} presents the outcomes in terms of the word error rate (WER) from fine-tuning Whisper-Tiny using diverse strategies including full-parameter fine-tuning, partial-parameter fine-tuning and LoRA. When fine-tuning only the decoder, we froze all weights in the audio encoder. Similarly, weights in the audio encoder and non-embedding layers in the text decoder were frozen when fine-tuning only the decoder embedding. As a preliminary experiment, greedy decoding was used, leading to a higher WER in the zero-shot experiment compared to that reported in \cite{radford2022robust}. 
    We observe evident enhancements in model performance across all fine-tuning experiments on both FLEURS and CommonVoice 9. In agreement with \cite{rouditchenko2023comparison}, our findings demonstrate that fine-tuning both the audio encoder and text decoder maximises performance. However, it is surprising that applying high-rank LoRA leads to the greatest model improvement on the FLEURS test set, with less than half the number of trainable parameters compared to full-parameter fine-tuning with decoupling embedding. The gain in model performance is reduced when a lower rank is used in LoRA.
        \begin{table*}[h]
    \centering
    \caption{\textit{WER (\%) results on FLEURS with various Whisper model size fine-tuning on the FLEURS training set. We reported results from greedy, beam search and our Filter-Ends + Min Lookahead = 3 decoding algorithms.}}
    \begin{tabular}{l c c c c c c c c c c c c} 
        \toprule
        Whisper Model & \multicolumn{3}{c}{Zero-shot} & \multicolumn{3}{c}{Full-parameter} & \multicolumn{3}{p{3cm}}{Decoupling embedding} & \multicolumn{3}{c}{LoRA} \\ [0.5ex] 
        \midrule
        {} & Greedy & \begin{tabular}{@{}c@{}}Beam \\ search\end{tabular} & Ours & Greedy & \begin{tabular}{@{}c@{}}Beam \\ search\end{tabular} & Ours & Greedy & \begin{tabular}{@{}c@{}}Beam \\ search\end{tabular} & Ours & Greedy & \begin{tabular}{@{}c@{}}Beam \\ search\end{tabular} & Ours \\
        Tiny & 74.78  & 58.17 & 56.74 & 41.38 & 39.30 & \textbf{38.77} & 41.54 & 39.84 & 39.42 &  42.89 & 40.59 &  39.96\\
        Base & 46.52 & 41.00 & 39.24 & 30.11 & 29.13 & \textbf{29.11} & 31.25 & 30.00 & 29.81 & 32.76 & 30.78 & 30.38\\
        Small & 23.34 & 21.96 & 21.25 & 18.51 & 17.86 & 17.86 & 18.52 & 17.85 & \textbf{17.84} & 19.55 & 18.77 & 18.76 \\
        Small \cite{rouditchenko2023comparison} & 22.52 & - & - & 19.03 & -& - & -& - & - & - & - & -\\
        Medium & 14.09 & 13.11 & 12.91 & 13.46 & 12.28 & \textbf{12.23} & 13.14 & 12.28 & 12.26 & 13.74 & 13.03 & 12.98 \\
        % Medium & 14.09 & & 12.91 & 14.71 & & 13.53 & 14.66 & & 13.94 & 13.02 & & \textbf{12.43} \\
        Medium \cite{rouditchenko2023comparison} & 13.46 & - & - & 12.38 & - & - & - & - & - & - & - & -\\
        Large-v2 & 12.04 & 11.09 & 10.55 & 10.57 & 9.84 & 9.82 & 10.18 & 9.49 & \textbf{9.44} & 10.58 & 9.69 & 9.62\\
        \bottomrule
    \end{tabular}
    \label{table:finetuning-other}
    \end{table*}
    In addition, enabling the independent learning of token input and output embeddings yields an average WER improvement of 1.18 on FLEURS and CommonVoice 9 upon full-parameter fine-tuning of Whisper-Tiny. This effect is even more pronounced with an average WER reduction of 2.64 when fine-tuning only the text decoder token embedding. 
\subsection{Decoding algorithms Whisper-Tiny}
    We chose 11 languages with a range of training dataset sizes and reported WERs \cite{radford2022robust}. Note that we did not perform experiments on some low-resource languages where WERs of over 90 were reported, since the model mostly produced incomprehensible output that a change in algorithm would not be able to significantly improve. 
    
    Firstly, the model was tested with a standard beam search with beam width 5. The WERs do not exactly match the values reported by OpenAI, but on average, they are similar. We compare to our values rather than the reported values in later experiments since all our experiments were run on the same system. Next, Filter-Ends was tested, and we observe a significant reduction of WER, particularly in the low-resource languages. Detailed results can be found in Appendix \ref{appendix:Additional Results}. 
    
    Using Filter-Ends differs from other changes to the decoding algorithm in that it does not increase the runtime. With Filter-Ends still in use, three modifications to the algorithm that increase runtime were tested: increased beam width, basic lookahead and min lookahead. We used larger beam widths up to a maximum of 20 to give a fair comparison of the lookahead algorithms, since they all increase the runtime. The min lookahead algorithm decreases WER by significantly more than either the increased beam width or basic lookahead, so we conclude that min lookahead gives better performance than is possible by the other methods.
\subsection{Decoding algorithms other Whisper models}

    We question whether the same encouraging results on Whisper-Tiny can be replicated with larger models, with the results in Table \ref{table:decoding-algorithm-all}. Due to the computation time of these larger models, the tests were only run with a standard beam search, Filter-Ends only and the best-performing configuration on Whisper-Tiny (Filter-Ends and Min Lookahead = 3).

    On all models, our algorithm outperforms the standard beam search on average across all languages. On larger models, we observe that low-resource languages benefit much more from our algorithms than high-resource languages. The smaller improvements on high-resource languages can be attributed to a lack of room for improvement on models that are already successful on those languages.
    
\subsection{Fine-tuning other Whisper models}
     To further examine the benefits of fine-tuning on larger-scale Whisper models, we extended our investigation to include the three best fine-tuning strategies: full-parameter, full-parameter with decoupling embeddings and LoRA on all Whisper models. We set LoRA rank $r$ to be half the number of attention dimensions in Whisper-Base, Whisper-Small, and a quarter of that in Whisper-Medium and Whisper-Large.
     
     To establish benchmarks, we used 10 hours of Vietnamese speech in the FLEURS training set, with the intent of incorporating the VinBigdata-VLSP2020-100h dataset in future studies. Outputs from greedy decoding, standard beam search with beam width 5 and our advanced lookahead beam search are listed in Table \ref{table:finetuning-other}. We used a learning rate of $10^{-5}$ to fine-tune Whisper-Base and Whisper-Small but $10^{-6}$ for Whisper-Medium and Whisper-Large. Other details for these experiments are listed in Appendix \ref{appendix: hyperparameters}. 
     
     All fine-tuning experiments achieve better WERs in Whisper-Base and Whisper-Small with an average WER improvement of 15.15 and 4.48 respectively when using greedy decoding. However, the benefits from fine-tuning Whisper become less significant as the model size increases. Interestingly, the gap in performance improvement between full-parameter fine-tuning and LoRA decreases as the model size grows. In Whisper-Medium, applying our decoding algorithm upon full-parameter fine-tuning achieves the best improvement, surpassing LoRA by 0.75. However, this difference reduces to 0.2 in Whisper-Large. Overall, the best improvement in each Whisper model is observed when applying Filter-Ends and Min Lookahead $= 3$ on top of fine-tuning.
\section{Conclusion}
     Despite having less trainable parameters, fine-tuning Whisper-Medium and Whisper-Large with high-rank LoRA yields comparable performance improvements in comparison to full-parameter fine-tuning. The decoupling of input and output embeddings does not harm the model performance but can occasionally surpass the results achieved through full-parameter fine-tuning. Furthermore, we suggest Filter-Ends and Min Lookahead as improvements to Whisper's decoding algorithm. We prove that Min Lookahead is expected to outperform standard beam search, while empirical results verify this with particularly strong performance on low-resource languages. Future studies should perform fine-tuning experiments on more low-resource languages and investigate increasing the beam's diversity as a potential improvement to the decoding algorithm.
\label{sec:conclusion}

\bibliographystyle{IEEEbib}
\bibliography{main.bib}

\clearpage
\onecolumn
\begin{appendices}
\section{Proof of Better Expected Results with Min Lookahead}
\label{appendix: min lookahead proof}

    \subsection{Definitions}

        A lookahead tree with lookahead \(m\), beam width \(n\) and base sequence \(b\) consists of the \(n\) greatest probabilities outputted by the decoder for some possible base sequences. The inputted sequences are \(b\) and for each \(1\leq k < m\), the concatenation of the previously inputted sequence with the token that has the greatest probability in the previous output of the decoder. Note that this lookahead tree is an element of \([0\,1]^{mn+1}\).
        
        A lookahead algorithm is a function \(f_m: \mathbb [0\,1]^{mn+1}\times\mathbb [0\,1]^{mn+1}\to \{\mathrm{False}, \mathrm{True}\}\) that uses the probabilities in the lookahead trees of two base sequences to choose one of them by outputting $\mathrm{True}$ to choose the first sequence and $\mathrm{False}$ to choose the second sequence.
        
        Let \(A,B\) be base sequences. Let \(A\) have probability \(x\) and \(B\) have probability \(y\).
        
        For all \(1\leq i \leq m\) and \(1\leq j \leq n\), let \(a_{ij},b_{ij}\) be the probability outputted by the decoder for the \(j\)-th most likely token in the \(i\)-th lookahead step for each of \(A,B\) respectively. These values are normalised so that for all $1\leq i \leq m$,
        
        Let the probability that the token with probability \(a_{ij}\) is in the correct sequence with base \(A\) be
        \begin{equation}
        p_{ij}=\frac{a_{ij}}{\sum_{l=1}^n a_{il}}\prod_{k=1}^{i-1}\frac{a_{k1}}{\sum_{l=1}^n a_{kl}}
        \end{equation}
        
        Let the probability that the token with probability $b_{ij}$ is in the correct sequence with base \(B\) be

        \begin{equation}
        q_{ij}=\frac{b_{ij}}{\sum_{l=1}^n b_{il}}\prod_{k=1}^{i-1}\frac{b_{k1}}{\sum_{l=1}^n b_{kl}}
        \end{equation}
        
        See Fig. 1 for the lookahead tree of \(A\).

\begin{figure}[h]
    \centering
    \includegraphics[width=0.5\linewidth]{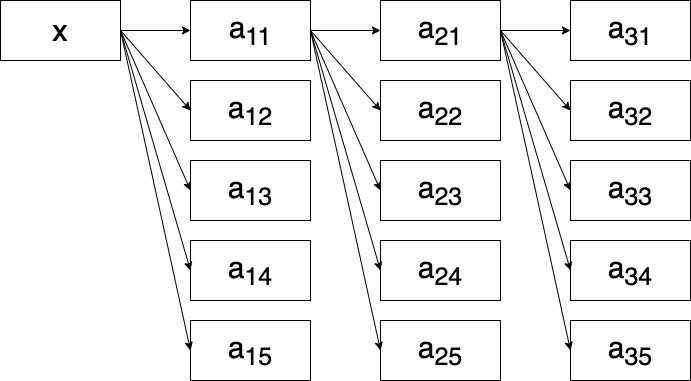}
    \caption{Example Lookahead tree with base sequence A, beam width \(n=5\) and lookahead \(m=3\)}
    \label{fig:enter-label}
\end{figure}

        The min algorithm is the lookahead algorithm \(\beta_m\) that outputs \(\beta_m(A,B)= \mathrm{True}\), if and only if the following inequality holds

        \begin{equation}
            \log(x)-\log(y)+ \sum_{i=1}^m\sum_{j=1}^n \mathrm{min}(xp_{ij}, yq_{ij})(\log(a_{ij})-\log(b_{ij}))>0,
        \end{equation}
        
        Note that \(\beta_0\) is the standard beam search algorithm.
        
        Given the score function of the beam search is a sum of logprobs, we use \(S(X)\) as a score function of a sequence \(X\) where \(S(X)\) is also the sum of the logprobs of tokens in \(X\).
        
        The expected score after \(m\) steps, \(E_m(X)\), of a base sequence \(X\) is the expected value of \(S(X^{\prime})\) across all \(X^{\prime}\), where $X^{\prime}$ is a concatenation of the base sequence \(X\) and a sequence of length \(m\). Note that we only consider sequences \(X^{\prime}\) with each token in the top \(n\) possible, since these are the only sequences that can possibly result from a lookahead algorithm. For each \(1\leq i\leq m\), let \(\alpha_i\) be the ranking in probability of a token at lookahead step $i$. Given the previous sequence of tokens ranked \(\alpha_1,\alpha_2,\dots,\alpha_{i-1}\), let \(\phi_{\alpha_1\alpha_2\dots\alpha_i}\) be the probability of the token ranked \(\alpha_i\)-th most likely in the \(i\)-th lookahead step. Thus,
        
        \begin{equation}
            E_m(X)=\log(x)+\sum_{j_1=1}^n \frac{\phi_{j_1}}{\sum_{l=1}^n \phi_{l}}(\log(\phi_{j_1})+\sum_{j_2=1}^n \frac{\phi_{j_1j_2}}{\sum_{l=1}^n \phi_{j_1l}}(\log(\phi_{j_1j_2})
            +\dots+\sum_{j_m=1}^n \frac{\phi_{j_1j_2\dots j_m}}{\sum_{l=1}^n \phi_{j_1j_2\dots j_{m-1}l}}\log(\phi_{j_1j_2\dots j_m})\dots))
        \end{equation}

        \begin{theorem}
        If \(\beta_m(A,B)=\mathrm{True}\) and \(\beta_0(A,B)=\mathrm{False}\), then \(E_m(A)-E_m(B)\) is most likely non-negative. Similarly, if \(\beta_m(A,B)=\mathrm{False}\) and \(\beta_0(A,B)=\mathrm{True}\), then \(E_m(A)-E_m(B)\) is most likely non-positive. Thus, \(\beta_m\) is expected to outperform \(\beta_0\).
        \end{theorem}

        \begin{proof}
        Consider \(\beta_m(A,B)\) and \(\beta_0(A,B)\) as estimates for whether \(E_m(A)-E_m(B)>0\).
        
        The sum used in the calculation of \(\beta_m\) is a weighted sum of the terms \(\log(\phi_{j_1j_2\dots j_k})\) in \(E_m(A)\). The weightings of these terms in \(E_m(A)\) are equal to \(xp_{ij}\), which is at least \(\min(xp_{ij}, yq_{ij})\).
        
        The leftover weightings of the terms we have partially considered are of the form \((xp_{ij}-yq_{ij})\log(a_{ij})\) and \((yp_{ij}-xp_{ij})\log(b_{ij})\). Noting that \(\beta_m(A,B)=\mathrm{True}\) and \(\beta_0(A,B)=\mathrm{False}\), it is most likely that \(a_{ij}\) terms are at least the \(b_{ij}\) terms.
        
        It is also expected that the logprobs not used in either estimate are on average distributed in the same way as the known logprobs. As the total weightings not considered are the same for both \(E_m(A)\) and \(E_m(B)\), we thus expect the unconsidered scores in \(E_m(A)-E_m(B)\) to be non-negative.
        
        Since \(\beta_m(A,B)=\mathrm{True}\), the considered scores are also non-negative, so the expected value of this estimate of \(E_m(A)-E_m(B)\) is non-negative. Similarly, if \(\beta_m(A,B)=\mathrm{False}\) and \(\beta_0(A,B)=\mathrm{True}\), then \(E_m(A)-E_m(B)\) is most likely non-positive.
        \end{proof}

\section{Hyperparameters}
\label{appendix: hyperparameters}
Hyper-parameters for fine-tuning the Whisper models of various sizes are listed in Table \ref{table:fine-tuning-training-details}. All experiments were ran on an A40 48GB GPU. 
\begin{table}[h]
    \centering
    \caption{\textit{Hyper-parameters used to fine-tune Whisper models on FLEURS training set}}
    \label{table:fine-tuning-training-details}
    \begin{tabular}{lccccc} 
        \toprule
        & 
        \begin{tabular}{@{}c@{}}Batch \\ size\end{tabular} & 
        \begin{tabular}{@{}c@{}}Learning \\ rate\end{tabular} & 
        \begin{tabular}{@{}c@{}}LoRA \\ $r$\end{tabular} & 
        \begin{tabular}{@{}c@{}}LoRA \\ $\alpha$\end{tabular} & Epoch\\
        \midrule
        Tiny & 32 & $10^{-5}$ & 192 & 384 & 80\\
        Base & 16 & $10^{-5}$ & 256 & 512 & 50\\
        Small & 8 & $10^{-5}$ & 384 & 768 & 30\\
        Medium & 3 & $10^{-6}$ & 256 & 1024 & 10\\
        Large-v2 & 1 & $10^{-6}$ & 320 & 1280 & 2\\
        \bottomrule
    \end{tabular}
    \label{table:fine-tuning-details}
    \end{table}
\clearpage
\section{Additional Results}
\label{appendix:Additional Results}
Additional WER results on the FLEURS test set with different decoding algorithms on Whisper-Tiny on all tested languages are listed in Table \ref{table:decoding-algorithm-tiny}.
\begin{table}[!htbp]
    \centering
    \caption{\textit{WER (\%) results on the FLEURS test set with different decoding algorithms on Whisper-Tiny.}}
    \begin{tabular}{l *9c}
        \toprule
        \textbf{Decoding Algorithm} & \textbf{Eng\textsuperscript{1}} & \textbf{Spa\textsuperscript{1}} & \textbf{Fra\textsuperscript{2}} &  \textbf{Ind\textsuperscript{2}} & \textbf{Pol\textsuperscript{2}} & \textbf{Tur\textsuperscript{2}} & \textbf{Vie\textsuperscript{3}} & 
        \textbf{Hun\textsuperscript{3}} & \textbf{Improvement} \\
        \midrule
        OpenAI Value \cite{radford2022robust} & 12.4 & 15.9 & 41.4 & 51.7 & 45.6 & 42.5 & 60.0 & 83.8 & -0.01 \\
        \midrule
        Beam Search & 11.81 & 15.70 & 37.71 & 51.73 & 48.45 & 40.32 & 58.17 & 89.36 & 0.00 \\
        \midrule
        Basic Lookahead = 1 & 11.62 & 15.43 & 36.87 & 50.58 & 47.15 & 39.69 & 58.32 & 87.13 & 0.81 \\
        Basic Lookahead = 2 & 11.49 & 15.28 & 36.79 & 51.48 & 47.00 & 39.43 & 58.31 & 86.47 & 0.87 \\
        Basic Lookahead = 3 & 11.50 & 15.26 & 36.63 & 52.23 & 47.40 & 39.46 & 59.23 & 89.72 & 0.23 \\
        Basic Lookahead = 4 & 11.54 & 15.13 & 36.67 & 53.45 & 47.36 & 40.08 & 60.55 & 88.20 & 0.01 \\
        Basic Lookahead = 5 & 11.54 & 15.16 & 36.63 & 52.73 & 47.91 & 40.09 & 61.32 & 89.99 & -0.27 \\
        Basic Lookahead = 6 & 11.60 & 15.35 & 36.98 & 54.75 & 49.20 & 40.76 & 61.16 & 93.34 & -1.24 \\
        \midrule
        Filter-Ends (FE) & 11.86 & 15.70 & 37.56 & 51.40 & 46.94 & 40.07 & 57.91 & 84.31 & 0.94 \\
        \midrule
        Min Lookahead = 1 & 11.57 & 15.33 & 37.41 & 50.83 & 47.38 & 39.64 & 58.21 & 85.45 & 0.93 \\
        Min Lookahead = 2 & 11.52 & 15.25 & 36.74 & 51.29 & 47.07 & 39.04 & 58.52 & 87.16 & 0.83 \\
        Min Lookahead = 3 & 11.45 & 15.22 & 36.70 & 51.26 & 46.81 & 38.88 & 57.92 & 86.28 & 1.09 \\
        Min Lookahead = 4 & 11.53 & 15.20 & 36.46 & 50.95 & 45.96 & 38.80 & 58.34 & 86.20 & 1.22 \\
        Min Lookahead = 5 & 11.54 & 15.17 & 36.26 & 50.85 & 46.21 & 39.29 & 58.63 & 87.27 & 1.00 \\
        Min Lookahead = 6 & 11.53 & \textbf{15.12} & 36.10 & 50.96 & 45.75 & 39.10 & 58.29 & 86.17 & 1.28 \\
        \midrule
        FE + Beam Width = 8 & 11.55 & 15.63 & 37.03 & 51.49 & 46.12 & 39.61 & 57.96 & 83.35 & 1.32 \\
        FE + Beam Width = 10 & 11.53 & 15.46 & 37.04 & 51.36 & 45.63 & 39.29 & 58.01 & 82.97 & 1.49 \\
        FE + Beam Width = 12 & 11.42 & 15.35 & 36.93 & 51.70 & 45.37 & 39.32 & 58.37 & 82.79 & 1.50 \\
        FE + Beam Width = 15 & 11.54 & 15.30 & 36.90 & 51.36 & 45.30 & 39.10 & 58.35 & 83.03 & 1.55 \\
        FE + Beam Width = 20 & 11.48 & 15.29 & 36.80 & 51.89 & 45.02 & 39.23 & 58.18 & 82.94 & 1.55 \\
        \midrule
        FE + Basic Lookahead = 1 & 11.55 & 15.45 & 36.89 & 50.42 & 45.27 & 38.95 & 57.14 & 82.14 & 1.93 \\
        FE + Basic Lookahead = 2 & 11.47 & 15.32 & 36.90 & 51.66 & 44.79 & 39.13 & 57.17 & 81.72 & 1.89 \\
        FE + Basic Lookahead = 3 & 11.50 & 15.28 & 36.44 & 51.49 & 44.92 & 39.36 & 57.65 & 81.69 & 1.86 \\
        FE + Basic Lookahead = 4 & 11.56 & 15.19 & 36.35 & 52.14 & 44.60 & 39.31 & 57.80 & 81.77 & 1.82 \\
        FE + Basic Lookahead = 5 & 11.57 & 15.19 & 36.38 & 52.34 & 44.45 & 39.01 & 58.48 & 82.30 & 1.69 \\
        FE + Basic Lookahead = 6 & 11.68 & 15.40 & 36.63 & 52.53 & 44.99 & 39.28 & 58.70 & 82.63 & 1.43 \\
        \midrule
        FE + Min Lookahead = 1 & 11.55 & 15.36 & 37.46 & 50.72 & 44.85 & 38.89 & 57.38 & 81.70 & 1.92 \\
        FE + Min Lookahead = 2 & 11.47 & 15.30 & 36.76 & 50.95 & 44.83 & 38.56 & 57.16 & 81.66 & 2.07 \\
        FE + Min Lookahead = 3 & \textbf{11.39} & 15.24 & 36.66 & \textbf{50.20} & 44.46 & \textbf{38.39} & \textbf{56.74} & 82.07 & \textbf{2.26} \\
        FE + Min Lookahead = 4 & 11.45 & 15.29 & 36.39 & 50.78 & 44.32 & 38.58 & 57.06 & 81.92 & 2.18 \\
        FE + Min Lookahead = 5 & 11.48 & 15.21 & 36.21 & 50.58 & 44.46 & 38.79 & 57.20 & 81.66 & 2.21 \\
        FE + Min Lookahead = 6 & 11.48 & 15.17 & 36.19 & 51.23 & 44.34 & 38.80 & 57.18 & 81.84 & 2.13 \\
        \midrule
        FE + Mean Lookahead = 1 & 11.55 & 15.38 & 37.41 & 50.33 & 44.87 & 38.85 & 57.23 & 81.79 & 1.98 \\
        FE + Mean Lookahead = 2 & 11.49 & 15.29 & 36.96 & 50.51 & 44.65 & 38.81 & 56.98 & 81.70 & 2.11 \\
        FE + Mean Lookahead = 3 & 11.43 & 15.24 & 36.39 & 50.57 & 44.42 & 38.89 & 57.13 & 81.81 & 2.17 \\
        FE + Mean Lookahead = 4 & 11.52 & 15.27 & 36.23 & 50.70 & 44.29 & 39.01 & 57.21 & 81.78 & 2.16 \\
        FE + Mean Lookahead = 5 & 11.47 & 15.24 & 36.21 & 50.56 & 44.45 & 39.10 & 57.26 & 82.11 & 2.11 \\
        FE + Mean Lookahead = 6 & 11.45 & 15.23 & \textbf{35.96} & 51.51 & 44.44 & 39.20 & 57.71 & 82.56 & 1.90 \\
        \midrule
        FE + Max Lookahead = 1 & 11.55 & 15.36 & 37.46 & 51.12 & 44.85 & 38.87 & 57.37 & 81.88 & 1.85 \\
        FE + Max Lookahead = 2 & 11.47 & 15.31 & 36.73 & 51.04 & 44.49 & 38.73 & 57.31 & 81.48 & 2.08 \\
        FE + Max Lookahead = 3 & 11.45 & 15.27 & 36.36 & 50.42 & 44.31 & 38.83 & 57.03 & 82.11 & 2.18 \\
        FE + Max Lookahead = 4 & 11.48 & 15.25 & 36.34 & 50.93 & 44.23 & 39.07 & 57.32 & \textbf{81.44} & 2.15 \\
        FE + Max Lookahead = 5 & 11.47 & 15.13 & 36.15 & 51.39 & \textbf{44.17} & 39.10 & 57.65 & 81.45 & 2.09 \\
        FE + Max Lookahead = 6 & 11.45 & 15.19 & 36.12 & 52.00 & 44.53 & 39.17 & 57.78 & 81.62 & 1.92 \\
        \bottomrule
    \end{tabular}
    \label{table:decoding-algorithm-tiny}
    \end{table}
\end{appendices}
\end{document}